\documentclass[journal]{IEEEtran}

\ifCLASSINFOpdf
\else
  \usepackage[dvips]{graphicx}
\fi
\usepackage{url}
\usepackage{graphicx}
\usepackage{bm}
\usepackage{hyperref}
\usepackage{epsfig,amsmath,amssymb,verbatim,mathrsfs,array,layout,textcomp,latexsym,slashed,booktabs,color,mathtools,tikz}
\usepackage{siunitx}

\begin{document}
 \title{Template-free Pulse Height Estimation of Microcalorimeter Responses with PCA}

\author{To Chin Yu
\thanks{This work was supported by the U.S. Department of Energy under contract number DE-AC02-76SF00515. T.C. Yu is affiliated with the Department of Physics, Stanford University, Stanford, CA 94305, USA and SLAC National Accelerator Laboratory, 2575 Sand Hill Road, Menlo Park, CA 94025, USA (email: ytc@stanford.edu).}}

\maketitle

\begin{abstract}
We present a template-free method of estimating pulse height of micro-calorimeter signals based on principal component analysis (PCA). The method is shown to improve the resolution on a simulated dataset by 25\% compared to the standard filtering technique that uses the averaged pulse as an estimation for the signal template. This technique is useful for reconstruction of pulses in micro-calorimeters with a high degree of pulse-shape variation where finding the correct signal templates is difficult.

\end{abstract}

The standard technique for estimating pulse heights of microcalorimeter responses is optimal filtering (OF) which is basically a maximal-likelihood estimator. However, in order to apply optimal filtering one needs to first come up with a signal template. If the signal template does not match the shape of the true signal then the OF estimator will be biased. Since the shape of microcalorimeter pulses varies with the pulse height, this mismatch between template and true signal is unavoidable. It is then not obvious what is the optimal template to use in such a situation. In this paper we present a method for pulse height estimation that does not require a template as input. This template-free approach offers both convenience as well as improved performance over the OF method.

\section{Optimal Filtering}

In the OF formulation, we assume that the measured signal $y_i$ ($i=0...T$) is given by 
\begin{align}
    y_i=As_i+n_i
\end{align}
where $A$ is the height of the true signal, $s_i$ is the known signal template, and $n_i$ is some additive random noise. The subscripts denote instants in time.

In frequency domain this becomes
\begin{align}
    Y_k=AS_k+N_k
\end{align}

Here the subscripts denote frequency bins. It is assumed that the power spectral density (PSD) $J_k$ of the noise $N_k$ is known and stationary. The probability distribution of noise power $P(|N_k|)$ should be a Gaussian with mean 0 and variance $J_k$:
\begin{align}
    P(|N_k|) = \mathcal{N}(0,J_k)
\end{align}
which leads to the following likelihood function for the observed signal:
\begin{align}
    \mathcal{L} &= \prod_k P(|Y_k-AS_k|) \\ 
    &= \prod_k \frac{1}{\sqrt{2\pi J_k}}\exp\Big[-\frac{|Y_k-AS_k|^2}{2J_k}\Big] \\
    &= (const.) \exp\Big[-\frac{1}{2} \sum_k\frac{|Y_k-AS_k|^2}{J_k}\Big]
\end{align}

The maximum likelihood estimator for $A$ can thus be found by minimizing the following quantity w.r.t. $A$:
\begin{align}
    \chi^2(A) = \sum_k\frac{|Y_k-AS_k|^2}{J_k}
\end{align}
It is easy to show that the resulted estimator is the minimum-variance unbiased estimator via the Cram\`{e}r-Rao bound (CRB) \cite{kay1993fundamentals}. 

In reality the signal also has a unknown random time shift $t_0$:
\begin{align}
    y(t)=As(t+t_0)+n(t)
\end{align}
We can incorporate this effect by using a modified $\chi^2$:
\begin{align}
    \chi^2(A,t_0) = \sum_k\frac{|Y_k-AS_ke^{-i\omega_k t_0}|^2}{J_k} \label{eq:of_chi2}
\end{align}
We refer the reader to \cite{kurinsky2018low} for further details. Note that due to the non-linearity in $t_0$ the resulted estimator (which becomes a vector $ (\hat{A},\hat{t}_0) $) does not saturate the CRB and may no longer be the estimator with optimal resolution.

\section{PCA Filtering}

Principal Componenet Analysis (PCA) can be formulated as the minimization of the total reconstruction error \cite{hastie2005elements}. If we treat each time trace of response as a vector $\vec{y}=(y_0,y_1,...,y_T)$, the reconstruction error can be written as:
\begin{align}
    \chi^2 = \sum_j \Big|\vec{y}^{(j)}-\sum_m A_m^{(j)}\vec{s}_m\Big|^2 \label{eq:pca_chi2}
\end{align}
The upper index $(j)$ denotes the $j$-th trace in our dataset. $\vec{s}_m$ is the $m$-th principal component and $A_k^{(j)}$ are the corresponding coefficients. If we only have a single principal component, this reduces to 
\begin{align}
    \chi^2 = \sum_j \Big|\vec{y}^{(j)}- A^{(j)}\vec{s}\Big|^2
\end{align}
which is equivalent to the OF $\chi^2$ if the noise is white. PCA will also produce the template $\vec{s}$
from the data without the need to know the template beforehand, which is an attractive feature. The PCA algorithm is an example of template-free filtering.

Obviously applying this simple version of PCA directly to real data will lead to poor results since real noise is most often not white. Also, the random time shift $t_0$ is unaccounted for in this algorithm which further degrades its performance. In order to tackle these issues, we propose an enhanced version of the PCA algorithm with the following modifications:

\begin{enumerate}
    \item Perform the PCA in transformed frequency domain that linearizes the time shift
    \item Use the Expectation-Minimization (EM) PCA algorithm \cite{bailey2012principal} that includes weights
\end{enumerate}

We now examine these modifications in detail.

\subsection{Shift-invariant Frequency Domain}

A time shift is a highly non-linear operation in time-domain. However, in frequency domain it just amounts to a phase shift, as can be seen in equation \ref{eq:of_chi2}. This still cannot be turned into a PCA problem because the coefficient $Ae^{-i\omega_kt_0}$ is a function of the frequency while PCA coefficients  (equation \ref{eq:pca_chi2}) can only depend on the trace index $j$ and the latent variable $m$ but not $k$. 

We can fix this issue by replacing the phase in frequency domain with phase difference:
\begin{align}
    \phi_k \to \phi_k - \phi_{k-1}
\end{align}\
Then under a time shift, this becomes:
\begin{align}
    \phi_k+\omega_kt_0 &\to \phi_k -\phi_{k-1} + (\omega_k-\omega_{k-1})t_0 \\
    &= \phi_k -\phi_{k-1} + \Delta\omega t_0
\end{align}
Thus under a time shift the coefficient now only picks up a frequency-independent factor $e^{-i\Delta\omega t_0}$ since the width of frequency bins $\Delta \omega$ is just a constant. The reconstruction error can be written in this transformed frequency domain as 
\begin{align}
    \chi^2 &= \sum_j \Big|\vec{Y}^{(j)}-\sum_m (A_m^{(j)}e^{-i\Delta\omega t_0^{(j)}})\vec{S}_m\Big|^2 \\
    &=\sum_j \Big|\vec{Y}^{(j)}-\sum_m B_m^{(j)}\vec{S}_m\Big|^2
\end{align}
where we define $B_m^{(j)} = A_m^{(j)}e^{-i\Delta\omega t_0^{(j)}}$. This problem can now be easily solved by complex-valued PCA.

\subsection{Expectation-Minimization (EM) Algorithm}

A non-white noise requires us to incorporate weights into the reconstruction error. This can be achieved using a method propose in \cite{bailey2012principal}. The basic premise is to apply the standard EM algorithm on the weighted reconstruction error which is essentially the likelihood function:

\begin{align}
    \mathcal{L}(\mathbf{B}|\mathbf{Y},\mathbf{S})=\chi^2 &= \vec{W}\cdot\sum_j \Big|\vec{Y}^{(j)}-\sum_m B_m^{(j)}\vec{S}_m\Big|_c^2 \\
    &= (\mathbf{Y}-\mathbf{B}\mathbf{S})\mathbf{W}(\mathbf{Y}-\mathbf{B}\mathbf{S})^\dagger
\end{align}
where the absolute value $|\cdot|_c^2$ is applied component-by-component. We also introduced the matrices $(\mathbf{Y})^j_k=Y^{(j)}_k$, $(\mathbf{B})^j_m=B^{(j)}_m$, $(\mathbf{S})^m_k=S^{m}_k$ and $(\mathbf{W})^k_\ell=(1/J_k)\delta_{k\ell}$.

More details of this algorithm can be found in  \cite{bailey2012principal}. Below we work out explicitly the update equations for the E-step and the M-step.

For the E-step, we fix the templates $\mathbf{S}$ and optimize the coefficients $\mathbf{B}$:
\begin{align}
    \frac{\partial \chi^2}{\partial {\vec{B}^{(j)}}} &=  (-\mathbf{S})\mathbf{W}(\vec{Y}^{(j)}-\vec{B}^{(j)}\mathbf{S})^\dagger   \\
     0 &=  \vec{Y}^{(j)}(\mathbf{S}\mathbf{W})^\dagger-\vec{B}^{(j)}\mathbf{S}(\mathbf{S}\mathbf{W})^\dagger
\end{align}
The last line is just a linear equation for $\vec{B}^{(j)}$.

We can obtain a similar equation for the M-step which we fix the coefficients and optimize the templates:
\begin{align}
    \frac{\partial \chi^2}{\partial {\vec{S}_m}} &=  \sum_j (-B^{(j)}_m)\mathbf{W}(\vec{Y}^{(j)}-\vec{B}^{(j)}\mathbf{S})^\dagger   \\
    0&=\sum_j \vec{Y}^{(j)}B^{*(j)}_m\mathbf{W}-\sum_j\vec{B}^{(j)}\mathbf{S}B^{*(j)}_m\mathbf{W} \\
    \sum_j \vec{Y}^{(j)}B^{*(j)}_m\mathbf{W}&=\sum_n\vec{S}_n\sum_j B^{(j)}_n B^{*(j)}_m\mathbf{W}
\end{align}
By concatenating the equations for different values of $m$ we will obtain a very large system of equations. Alternatively we can consider only a single template at a time:
\begin{align}
    \sum_j \vec{Y}^{(j)}B^{*(j)}_n\mathbf{W}=\vec{S}_n\sum_j |B^{(j)}_n|^2 \mathbf{W} \label{eq:Mstep}
\end{align}

After each template is obtained we subtract the corresponding component from the data and find the next template. In our trials, we uses equation \ref{eq:Mstep} without data subtraction for better speed performance. This implicitly assumes that the off-diagonal terms in $\chi^2$ are sub-dominant, i.e. the matrix $\mathcal{B}_{mn}=\sum_j B^{*(j)}_m B^{(j)}_n$ is diagonally dominant, which usually means that the final templates obtained could be rotated with respect to the ones obtained from solving the full system. The resultant performance in terms of energy resolution usually does not suffer much. At the end of this step the templates are smoothed and re-orthonormalized.

The optimality of this algorithm is guaranteed by the CRB since it is a linear model under the assumption that the true signal actually lies within the appropriate vector space, and that the EM procedure correctly converges to the true minima. Note that in this case, the CRB guarantees minimum variance only in the amplitude but not the phase. In other words we attain optimal resolution in pulse height but not necessarily time shift.

\section{Numerical Study}

\begin{figure}
    \centering
    \includegraphics[width=0.4\textwidth]{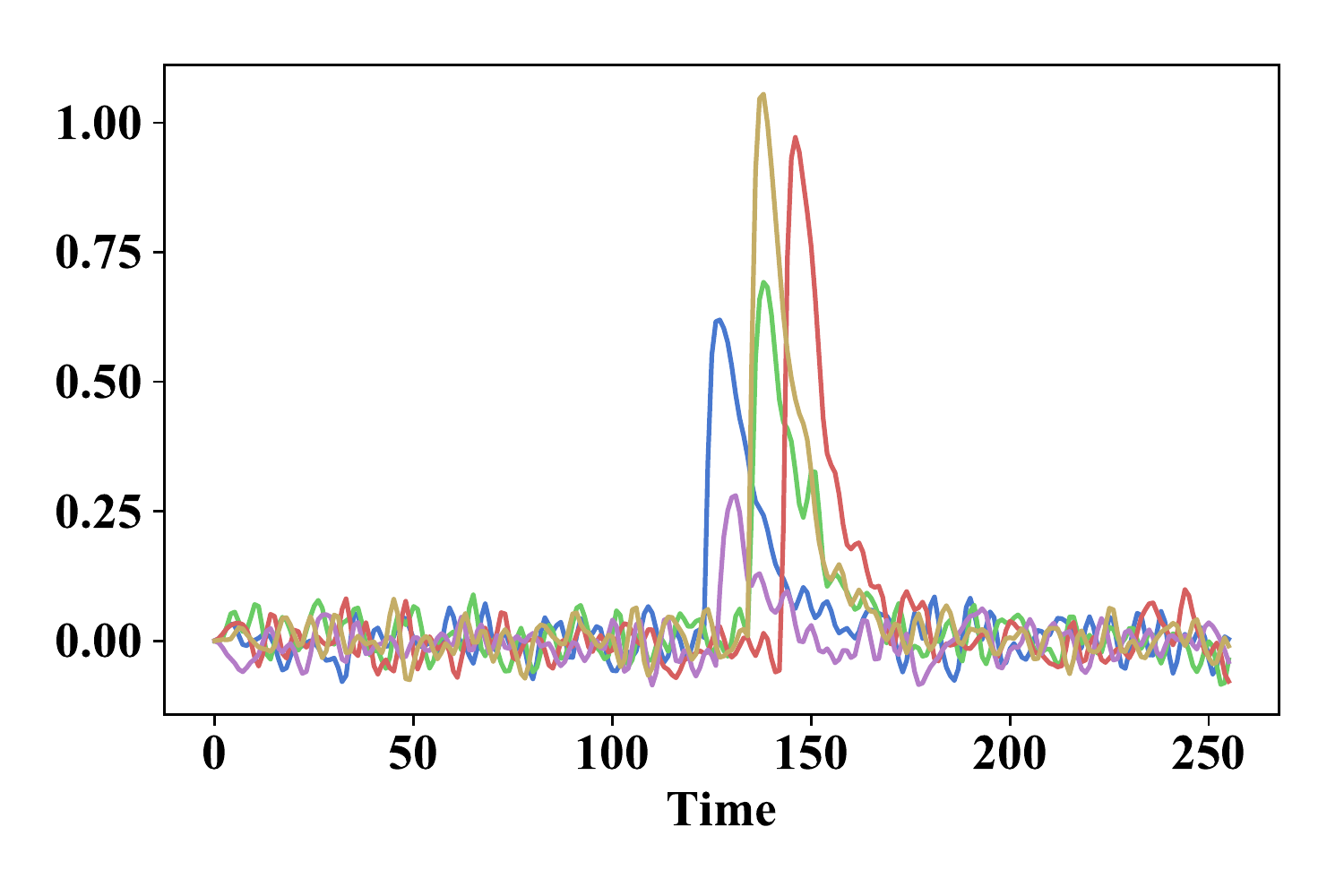}
    \caption{Examples of simulated pulses.}
    \label{fig:toy_model}
\end{figure}

We tested our method and compared against OF using randomly generated pulses. The shape of the pulses is described by
\begin{align}
    y(t;\omega_1,\omega_2,t_0) =  e^{-\omega_1(t-t_0)}(1-e^{-\omega_2(t-t_0)}) \Theta(t-t_0)
\end{align}
where $t$ is an integer ranging from 0 to 255, the inverse time constants $\omega_1$ and $\omega_2$ are uniformly sampled from $[0.1,0.2]$ and $[0.5,0.6]$ respectively, the time shift $t_0$ is Gaussian sampled with mean $128$ (center of the trace) and variance $5$. We then scale each pulse so that it has an amplitude $A$ that is uniformly drawn from $[0.1,0.9]$.  We used a sample of 1000 traces as our test dataset. On top of the pulse we also overlay low frequency noise at $5\%$ level with a cut-off frequency $f_{cutoff}$ of $0.2$ units (Fig. \ref{fig:toy_model}). 

\begin{figure}
    \centering
    \includegraphics[width=0.45\textwidth]{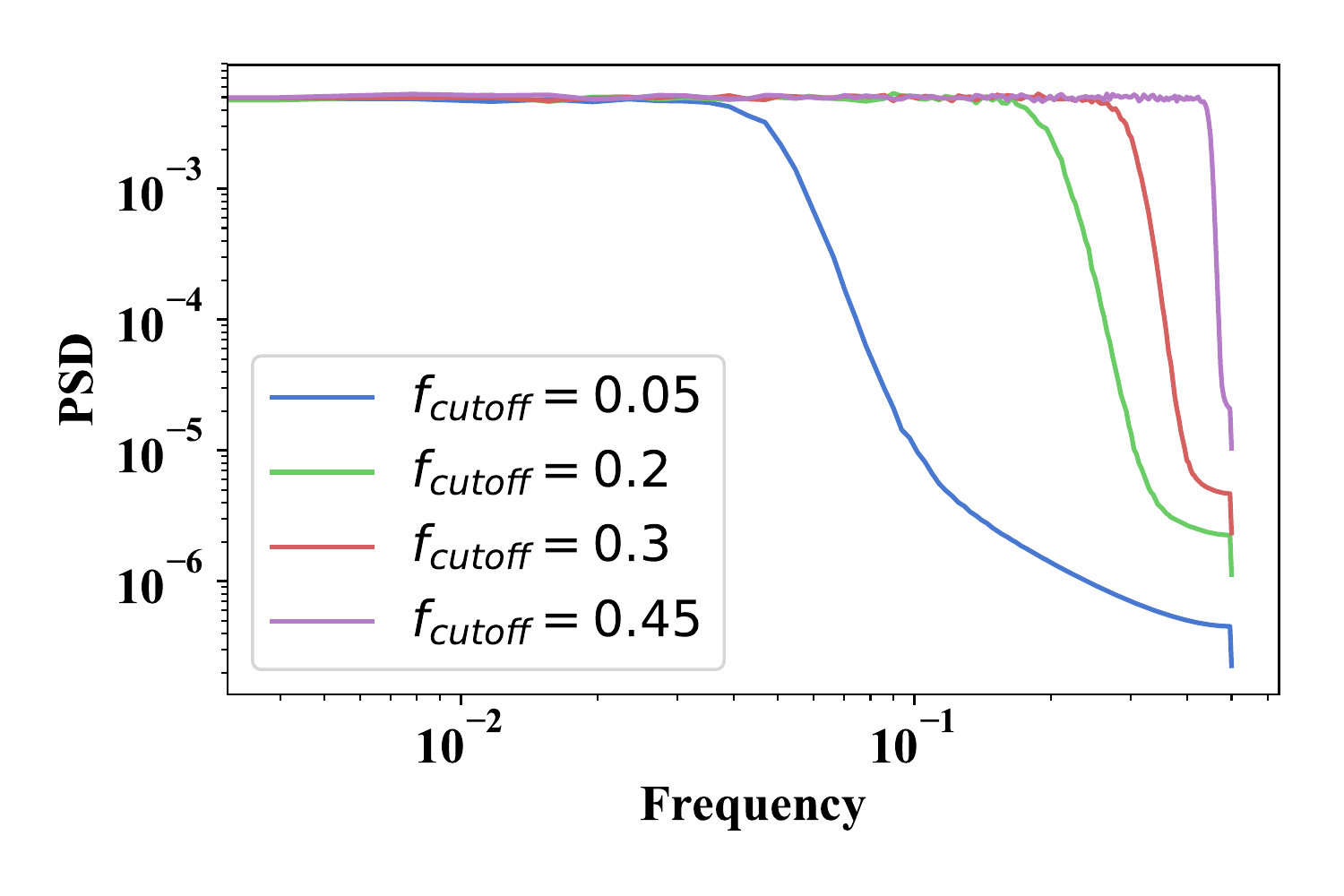}
    \caption{The noise PSD with various $f_{cutoff}$. We used the value $f_{cutoff}=0.2$ for our main simulated dataset.}
    \label{fig:PSD}
\end{figure}

We used the summed norm of the PCA coefficients as the pulse height estimator. We used 2 PCA components. We performed the standard OF with time shift to obtain the OF estimator. Both the PCA and the OF estimators are calibrated against true amplitude using a linear fit before comparison. We also included the pulse integral as a baseline estimator.

\begin{figure}
    \centering
    \includegraphics[width=0.4\textwidth]{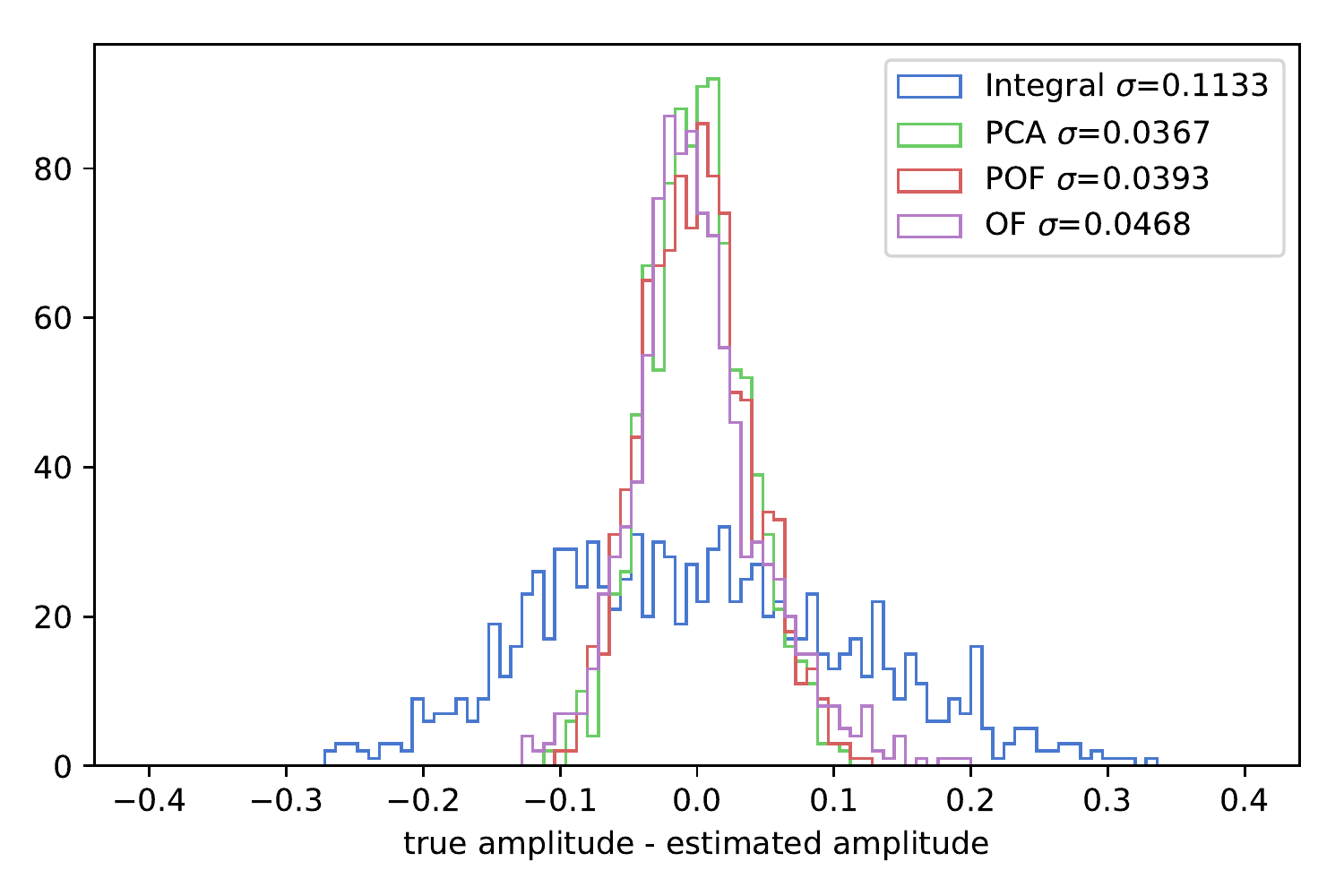}
    \caption{Comparison of pulse height resolutions between simple integral ($\sigma=0.11$), PCA estimator ($\sigma=0.0367$), PCA-assisted optimal filter (POF) amplitude ($\sigma=0.0393$) and optimal filter amplitude ($\sigma=0.0469$). }
    \label{fig:toy_model_res}
\end{figure}

In addition, we tried using the first principal component as the OF template instead of the mean pulse. We refer to this as PCA-assisted OF (POF). The first principal component can contain an overall phase factor which has to be determined by hand so this method requires some manual tuning to pick the useful template. Done correctly this leads to an improvement in resolution (Fig. \ref{fig:toy_model_res}).

\section{Hyperparameter tuning}

The optimality of the algorithm is only guaranteed when the global minima is attained during the iteration. As in most high-dimensional optimization problems, the loss landscape is often very jagged which makes it difficult to converge to the global minima. In practice, various regularization techniques are used to smooth the loss landscape \cite{li2018visualizing}. 

In our case, the regularization is effectively performed by smoothing of the template. The smoothing algorithm we used is the Savitzky-Golay filter\cite{savitzky1964smoothing} which generates three additional hyperparameters that we can tune - the filter window $w$, the polynomial order $p$, and the order of derivative used $d$. Intuitively a large window and low polynomial order will lead to a stronger regularization. In total there are 4 tunable hyperparameters including the number of principal components used.

\begin{figure}
    \centering
    \includegraphics[width=0.42\textwidth]{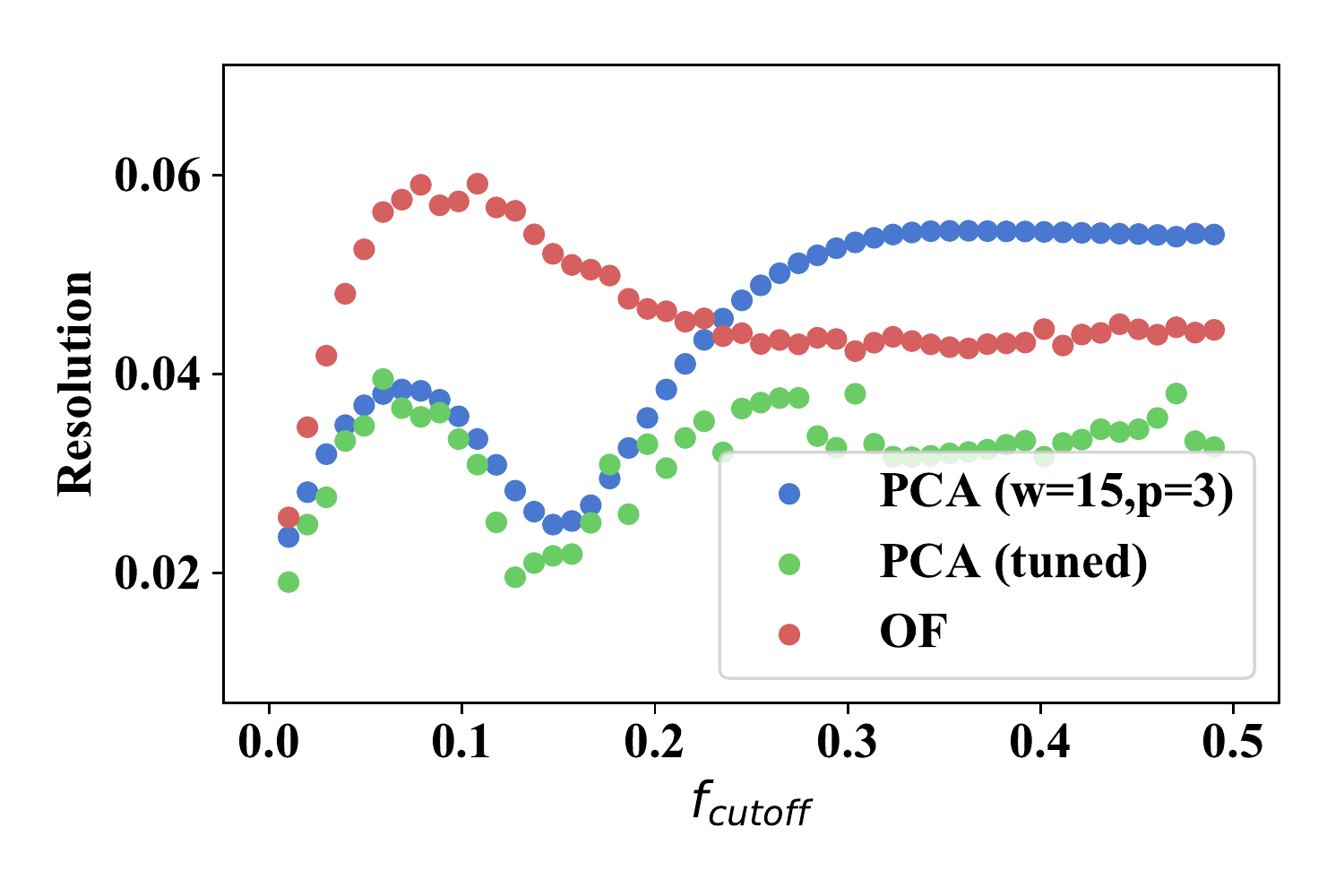}
    \caption{Comparison of pulse height resolutions between PCA with fixed hyperparameters, PCA with tuned hyperparamters, and optimal filter.}
    \label{fig:res_with_tuning}
\end{figure}

Hyperparameter tuning is important for the correct convergence, and thus the performance, of the algorithm. We performed some simple scans across datasets with different values of $f_{cutoff}$ in the noise PSD and studies the effect of some of the hyperparameters (Fig. \ref{fig:res_with_tuning}). A more complete survey will be left to future work.

\section*{Acknowledgment}
The author would like to thank Noah Kurinsky, Nicholas Mast, Emanuele Michielin and Jonathan Wilson for useful discussions and feedback.

\nocite{*}
\bibliographystyle{plain}
\bibliography{refs.bib}

\end{document}